\documentclass[
showkeys,	
reprint,	
amsmath,	
amssymb,	
amsfonts,	
aps,		
prl%
groupedaddress,
superscriptaddress,
a4paper,%
nofootinbib,
eprint,		
final,		
preprintnumbers, 
]{revtex4-2}

\usepackage[tbtags]{mathtools}
\usepackage{amsmath,amsfonts,mathdots,amssymb,yfonts,calc}
\usepackage{graphicx,graphics,xcolor}
\usepackage{subfigure}
\usepackage{amsthm}

\usepackage{natbib}
\usepackage{booktabs}
\usepackage{siunitx}
\usepackage{float}
\usepackage{tabularray}
\usepackage{multirow}
\usepackage{physics}
\usepackage{dcolumn}
\usepackage[utf8]{inputenc}  
\usepackage[T1]{fontenc} 
\usepackage{natbib}
\usepackage{booktabs}
\usepackage{hyperref}
\hypersetup{
    colorlinks=true,
    linkcolor=blue,
    filecolor=magenta,      
    urlcolor=blue!80,
    citecolor=cyan,
    linktocpage=true,
    breaklinks=false,
}

\begin{document}
\title{Breakdown of Linear Response in Uniformly Hyperbolic Systems with Hierarchical Structure}
\author{Vinesh Vijayan}
\email[]{vinesh.physics@rathinam.in}
\affiliation{Department of Science and Humanities, Rathinam Technical Campus, Coimbatore, India -641021}
\author{Priyadharshini B}
\affiliation{Department of Science and Humanities, Rathinam Technical Campus, Coimbatore, India -641021}
\author{Santhoshbalaji M}
\affiliation{Department of Science and Humanities, Rathinam Technical Campus, Coimbatore, India -641021}
\author{Mohanasundari M}
\affiliation{Department of Science and Humanities, Rathinam Technical Campus, Coimbatore, India -641021}
\date{\today}

\begin{abstract}
Linear response theory asserts that sufficiently small external biases produce currents proportional to the applied force and forms the theoretical foundation of nonequilibrium transport. Here we demonstrate that linear response can break down even in uniformly hyperbolic deterministic systems when hierarchical asymmetry is present. Using a minimal class of uniformly expanding chaotic maps with hierarchical multiscale structure, we show that progressively finer transport channels become dynamically active as the applied bias decreases. The resulting force–current relation is monotone and exhibits a hierarchical, fractal-like organization of activation thresholds. As a consequence, the effective mobility diverges as $F \to 0$, demonstrating breakdown of linear response despite strong chaos and uniform hyperbolicity. The effect arises from deterministic multiscale activation rather than intermittency, stochastic noise, or singular invariant measures. These results identify hierarchy as an independent deterministic mechanism for nonperturbative transport response and demonstrate that uniform hyperbolicity alone does not guarantee the validity of linear response.
\end{abstract}

\maketitle
\pagestyle{plain}

\section{\label{s1}Introduction}
Linear response theory predicts that sufficiently small external forces generate currents proportional to the applied bias, providing the theoretical foundation of transport in nonequilibrium statistical physics \cite{Kubo1957,Kubo1991,Baiesi2009FluctuationsResponse}.
Rapid mixing is expected to suppress memory effects, extending validity to strongly chaotic systems\cite{Ruelle1989}. This expectation suggests that sufficiently strong chaos should stabilize linear response by rapidly decorrelating perturbations. Whether this intuition always holds in deterministic systems with multiscale structure remains an open question. Violations are typically associated with stochastic noise, intermittency, or marginally stable dynamics\cite{Gallavotti1995, Rondoni2007, Dorfman1999}. In this Letter, we show that none of these ingredients are necessary. A purely deterministic chaotic system with hierarchical structure destroys linear response. The mechanism is neither stochastic nor measure-theoretic, originating instead from activation of transport channels across infinitely many scales. Scale hierarchies appear from rough energy landscapes to multiscale geometries in physical systems\cite{Mandelbrot1982,Bouchaud1990}. Their role in purely deterministic transport remains poorly understood. Here we isolate hierarchy as an independent mechanism inducing nonperturbative response, even in uniformly hyperbolic systems.

Linear response theory was extended to chaotic systems by Ruelle and others, proving its validity in uniformly hyperbolic systems with smooth dependence of observables on external parameters \cite{Ruelle1999,Ruelle2009,Baladi2008LinearResponse}.
These results fostered the belief that chaos stabilizes rather than destroys linear response. Deterministic flows and maps have modeled noise-free transport, demonstrating anomalous diffusion and directed currents from asymmetry and chaos—yet assuming regular linear response at small biases despite nonlinearity at large forcing\cite{Geisel1982, Klages1995, Klages1997}.

Violations of linear response occur only under special conditions: intermittency/marginal stability\cite{Rondoni2007}, singular invariant measures, slow correlation decay\cite{Gallavotti1996}, or stochastic/thermostatted systems far from equilibrium\cite{Bouchaud1990}. These breakdowns are attributed to weak chaos, intermittency, or measure-theoretic pathologies. Hierarchical/multiscale structures appear in fractal diffusion, renormalization, and critical phenomena—typically as geometric decoration or sources of anomalous diffusion—while devil's staircase and Cantor set responses characterize depinning, hysteresis, and mode-locking\cite{Bak1986,Sethna2001}.

In this Letter, we show that linear response can fail even in uniformly expanding chaotic systems when hierarchical asymmetry is present, identifying multiscale activation as a deterministic mechanism for nonperturbative transport.

\section{\label{s2} Model}
Following standard constructions of chaotic transport models\cite{Geisel1982, Klages2007, Klages1995}, we consider a class of one-dimensional deterministic maps defined on the unit interval and lifted to the real line to allow transport:
\begin{equation}
x_{n+1} = T_F(x_n) := 2x_n + F + g_h(x_n),
\label{E1}
\end{equation}
where $F\in\mathbb{R}$ is a constant bias and $g_h(x)$ is a bounded asymmetric perturbation encoding the hierarchical structure. The reduced dynamics is obtained by taking $x_{n+1} \bmod 1$, and transport is quantified by the net displacement along the real line.
The function $g_h(x)$ is constructed as a convergent multiscale sum:
\begin{equation}
g_h(x) = \sum_{k=0}^\infty \epsilon^k g(2^k x \bmod 1), \quad 0 < \epsilon < 1
\label{E2}
\end{equation}
where the base function $g(x)$ is a piecewise constant asymmetric function on $[0,1)$ given by
\begin{equation}
g(x) = \begin{cases} 
0 & 0 \leq x < \frac{1}{2} \\ 
-\alpha & \frac{1}{2} \leq x < \frac{3}{4} \\ 
\beta & \frac{3}{4} \leq x < 1 
\end{cases}
\label{E3}
\end{equation}
with $\alpha, \beta > 0$. This function is bounded and piecewise constant with a dense set of discontinuities, and the series converges uniformly. The hierarchical sum generates analogous asymmetries at all finer spatial scales $2^{-k}$, with the choice of $g(x)$ introducing minimal left-right asymmetry at the base scale.
\subsection{\label{ss2}Uniform expansion and chaos}
\begin{figure*}[htbp!]
    \centering
    \includegraphics[width=0.80\linewidth]{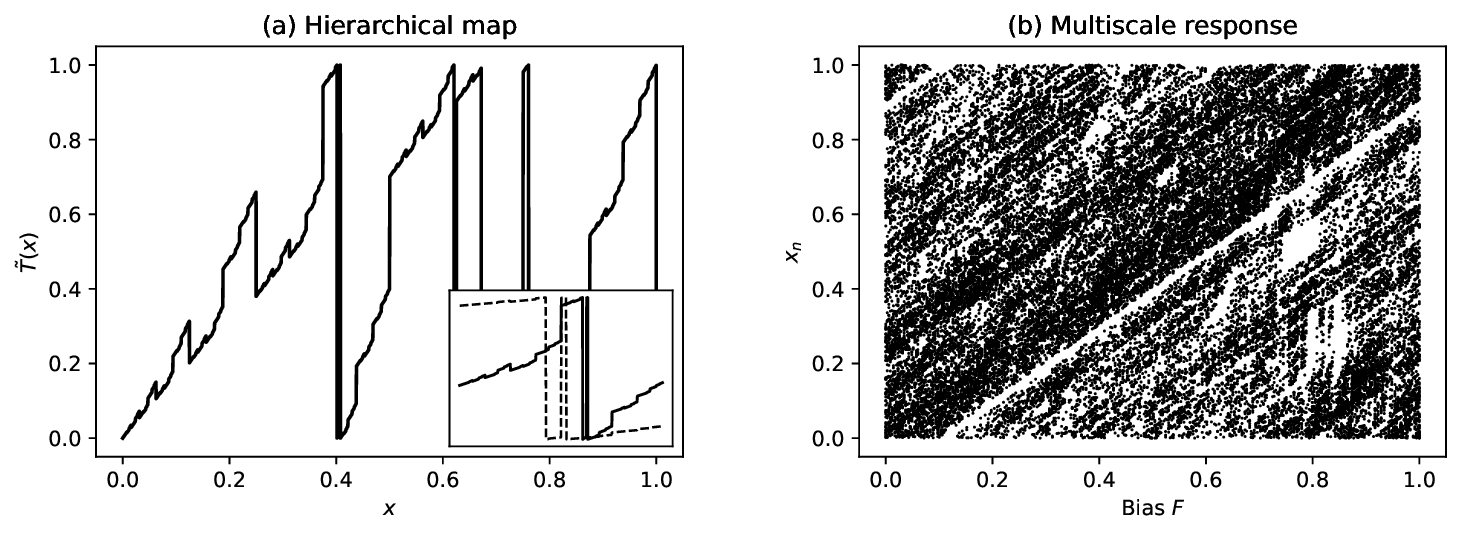}
    \caption{
    (a) Reduced hierarchical map $\tilde T(x)=T(x)\bmod 1$ showing a piecewise linear expanding structure with asymmetric discontinuities at all spatial scales.   The inset demonstrates self-similar collapse under magnification, revealing the hierarchical organization of the map. (b) Multiscale response of trajectories $x_n$ as a function of the applied bias $F$, illustrating strong sensitivity across a dense set of force values.
    }
    \label{fig1}
\end{figure*}
The map $T_F(x)$ is piecewise linear and differentiable away from discontinuities, with constant derivative
\begin{equation}
T_F'(x) = \frac{d}{dx} \bigl[2x + F + g_h(x)\bigr] = 2 \quad \text{for almost every $x$}.
\label{E4}
\end{equation}
The reduced dynamics modulo unity,
\begin{equation}
\tilde{T}_F(x) = T_F(x) \bmod 1,
\label{E5}
\end{equation}
is thus uniformly expanding, with positive Lyapunov exponent
\begin{equation}
\lambda = \int_0^1 \ln \bigl|\tilde{T}_F'(x)\bigr| \, dx = \ln 2,
\label{E6}
\end{equation}
independent of the hierarchical perturbation. The system therefore exhibits robust chaotic dynamics with exponential sensitivity to initial conditions. Importantly, the hierarchical perturbation alters neither the expansion rate nor the Lyapunov exponent, but only the structure of transport pathways\cite{Gaspard1998}. Uniform expansion guarantees the existence of an absolutely continuous invariant measure with bounded density, as established for piecewise expanding maps.

In the numerical simulations we use the parameters
$\alpha=0.3$, $\beta=0.6$, hierarchy amplitude
$\epsilon=0.4$, and hierarchy depth $L=8$.
FIG.~\ref{fig1}(a) demonstrates the hierarchical nature of the map. Beyond the base piecewise linear structure, additional asymmetric discontinuities appear at progressively finer spatial scales. The inset shows that upon magnification and rescaling, a local segment of the map reproduces the shape of a larger-scale segment, revealing explicit self-similarity. This recursive structure reflects the hierarchical construction of the perturbation and distinguishes the map from non-hierarchical chaotic systems. FIG.~\ref{fig1}(b) shows a multiscale response diagram, displaying chaotic trajectories as a function of the applied bias. Despite the absence of any bifurcation or change in dynamical stability, the trajectories reorganize across a dense set of force values, reflecting the hierarchical structure of the map.
\section{\label{s3}Deterministic Chaos Without Linear Response}

\begin{figure*}[htbp!]
\centering
\includegraphics[width=0.80\textwidth,keepaspectratio]{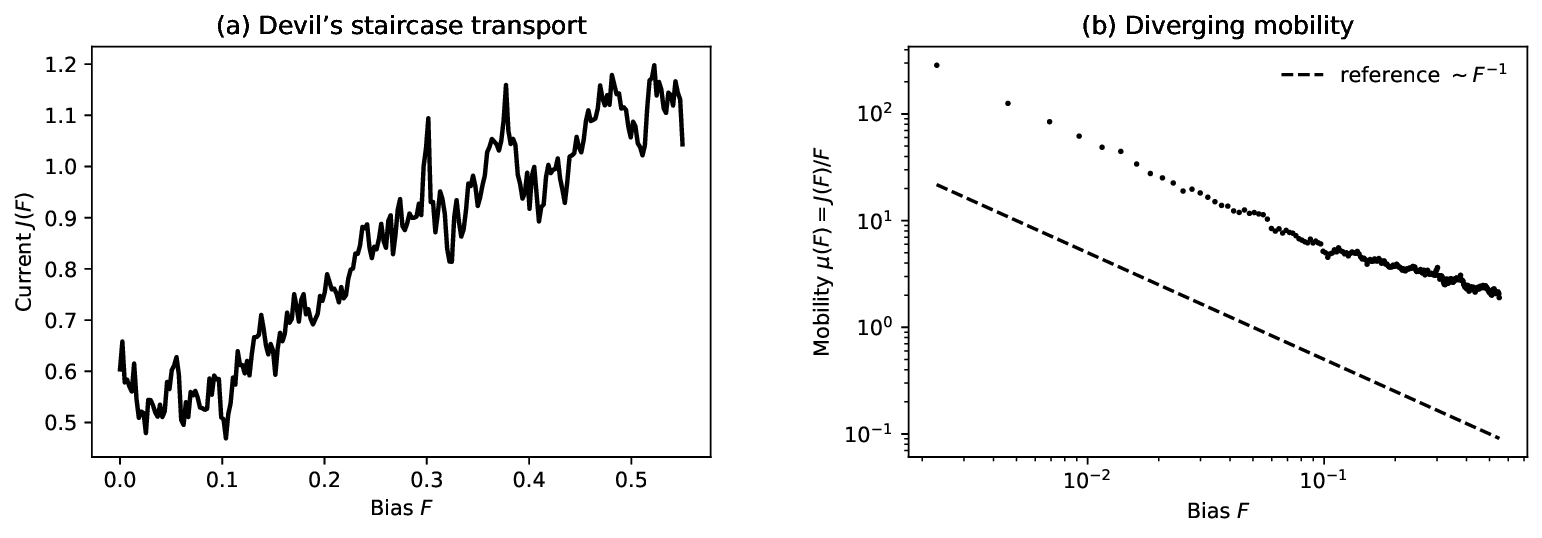}
\caption{
\textbf{(a)}~Devil's staircase current $J(F)$ exhibiting a monotone, fractal structure with hierarchical activation thresholds that accumulate as $F \to 0$.
\textbf{(b)}~(b) Log-log plot of mobility $\mu(F) = J(F)/F$ exhibiting unbounded growth as $F \to 0$, signaling breakdown of linear response. The dashed line indicates a reference slope $\sim F^{-1}$ shown for comparison only.
}
\label{fig2}
\end{figure*}
The transport current is defined as the average displacement per iteration,

\begin{equation}
J(F)=\lim_{n\to\infty}\frac{1}{n}\langle x_n-x_0\rangle ,
\label{E7}
\end{equation}

where the average is taken with respect to the invariant measure of the reduced dynamics.  
For standard chaotic systems with smooth parameter dependence, the current is expected to admit an analytic expansion near $F=0$,

\begin{equation}
J(F)=\mu F+O(F^2),
\label{E8}
\end{equation}

with finite mobility $\mu$. This relation expresses linear response.

In the present system this expansion fails because the hierarchical perturbation introduces asymmetric structures at arbitrarily fine spatial scales $2^{-k}$. Although the map remains uniformly expanding with constant slope $2$, backward iteration amplifies perturbations exponentially: after $k$ steps a bias $F$ produces an effective displacement of order $2^kF$. A hierarchical feature at scale $2^{-k}$ therefore becomes dynamically relevant when this amplified displacement becomes comparable to the spatial scale of the corresponding structure, yielding the activation condition

\begin{equation}
F_k \sim 2^{-k}.
\label{E9}
\end{equation}

As the bias decreases, progressively finer hierarchical components of the perturbation become dynamically resolved. The corresponding activation thresholds accumulate densely at $F=0$, implying that arbitrarily small biases can successively activate transport contributions from increasingly fine spatial scales.

To quantify this mechanism, we write

\begin{equation}
J(F)=F+\sum_{k=0}^{\infty}\epsilon^k I_k(F),
\label{E10}
\end{equation}

where

\begin{equation}
I_k(F)=\int g(2^k x)\rho_F(x)\,dx
\label{E11}
\end{equation}

and $\rho_F$ denotes the invariant density of the reduced map. Let

\begin{equation}
N(F)\sim\log_2(1/F)
\label{E12}
\end{equation}

denote the number of hierarchical levels whose activation thresholds satisfy $F_k\lesssim F$.

The hierarchical activation mechanism admits a quantitative estimate.

\textbf{Proposition (scale-by-scale lower bound).}  
There exists a constant $c_0>0$ such that for sufficiently small $F$ and for all hierarchical levels satisfying $a\le 2^kF\le b$ (for fixed constants $0<a<b<1$),

\begin{equation}
|I_k(F)|\ge c_0 .
\label{E13}
\end{equation}

The bound follows from the bounded density of the invariant measure and the discontinuous structure of the observable $g(2^k x)$. When $2^kF\sim O(1)$, the applied bias shifts trajectories across the discontinuity structure of this observable, producing an imbalance in its average that remains bounded away from zero independently of $k$.

This mechanism can be understood using standard mixing estimates for expanding maps. Observables of bounded variation exhibit exponential decay of correlations,

\begin{equation}
\begin{split}
\left|\int \phi(x)\psi(T^n x)\rho(x)\,dx
-\int \phi\rho\,dx \int \psi\rho\,dx\right| \\
\le C\,\|\phi\|_{BV}\|\psi\|_\infty \theta^n .
\end{split}
\label{E14}
\end{equation}

with $0<\theta<1$. Once $2^kF\sim O(1)$, the bias effectively shifts trajectories across the discontinuity of $g(2^k x)$, generating an $O(1)$ imbalance in the average of this observable. Exponential mixing ensures that fluctuations around this imbalance remain bounded, implying that the contribution $I_k(F)$ does not vanish as $k$ increases.

Summing the contributions of all activated scales yields

\begin{equation}
|J(F)|\ge c_0\sum_{k\le N(F)}\epsilon^k .
\label{E15}
\end{equation}

For $\epsilon=1$, this bound grows as $J(F)\gtrsim \log(1/F)$, while for $0<\epsilon<1$ it approaches a finite constant. Consequently,

\begin{equation}
\limsup_{F\to0}\frac{|J(F)|}{F}=\infty ,
\label{E16}
\end{equation}

and the effective mobility

\begin{equation}
\mu(F)=\frac{J(F)}{F}
\label{E17}
\end{equation}

diverges as $F\to0$. The current therefore lacks a finite linear-response coefficient.

The resulting force–current relation is monotone and organized by a dense hierarchy of activation thresholds. In chaotic systems these thresholds are broadened by dynamical fluctuations, producing a fractal response structure rather than an ideal piecewise-constant staircase. As the applied bias decreases, progressively deeper levels of the hierarchy become dynamically active, leading to an increasingly sensitive and nonperturbative transport response. FIG.~\ref{fig2}(a) shows the Devil’s staircase structure of the current, while panel (b) displays the diverging mobility $\mu(F)=J(F)/F$. For each bias value, the system was iterated for $3\times10^3$ transient steps followed by $8\times10^3$ iterations used for averaging. Bias values were sampled uniformly in the range $0\le F\le0.55$, and the results were averaged over 20 initial conditions.

\section{Discussion}

The breakdown of linear response identified here arises from the interplay between uniform expansion and hierarchical asymmetry. Unlike previously studied mechanisms for nonlinear response in chaotic systems, the present effect does not rely on intermittency, marginal stability, slow correlation decay, or singular invariant measures. The dynamics remains uniformly hyperbolic with a positive Lyapunov exponent for all parameters considered. Instead, the nonperturbative response originates from the activation of an unbounded hierarchy of transport channels whose characteristic thresholds accumulate at vanishing bias.

The essential mechanism is structural. Because the map expands uniformly, small perturbations are amplified exponentially under backward iteration. Consequently a bias $F$ resolves hierarchical features at progressively finer spatial scales. As $F$ decreases, deeper levels of the hierarchy become dynamically active, producing additional transport contributions that prevent the current from admitting a regular Taylor expansion near $F=0$.

This behavior leads to a force–current relation organized by a dense sequence of activation thresholds. In deterministic chaotic dynamics these thresholds are broadened by fluctuations associated with mixing, so that the response appears as a monotone fractal structure rather than an ideal piecewise-constant staircase. The absence of a finite linear-response coefficient reflects the fact that arbitrarily many hierarchical scales contribute as the bias approaches zero.

The effect is robust within the class of systems considered. Finite truncations of the hierarchy introduce a small-force cutoff determined by the deepest resolved scale but do not restore linear response within the accessible bias range. Weak smoothing or noise rounds individual activation features while preserving the multiscale structure responsible for the divergence of the effective mobility. The mechanism identified here does not depend on the specific form of the map used in our construction. The essential ingredients are uniform expansion together with a hierarchy of asymmetric structures whose characteristic scales accumulate geometrically. Whenever a small bias is exponentially amplified by chaotic dynamics while simultaneously resolving progressively finer asymmetric features, the number of activated transport channels increases as the bias decreases. This multiscale activation mechanism is therefore expected to arise in a broader class of deterministic systems possessing hierarchical structure.

These results highlight a fundamental limitation of linear response theory. Although uniform hyperbolicity ensures strong chaotic mixing, it does not guarantee smooth dependence of transport observables on external parameters when multiscale asymmetry is present.
\section{Conclusion}

We have investigated transport in a uniformly expanding chaotic map with hierarchical asymmetry and shown that linear response need not hold in such systems. The breakdown originates from the activation of progressively finer transport channels as the applied bias decreases. Because the number of dynamically relevant hierarchical levels grows as $F \to 0$, the effective mobility $\mu(F)=J(F)/F$ does not approach a finite limit.

The resulting force–current relation is monotone and organized by a dense hierarchy of activation thresholds, producing a fractal response structure rather than a regular expansion near zero bias. This mechanism does not rely on intermittency, weak chaos, or stochastic effects, but instead emerges from the interaction between uniform hyperbolicity and multiscale asymmetry.

More broadly, the results demonstrate that strong chaos alone does not guarantee the validity of linear response. Hierarchical organization can generate intrinsically nonperturbative transport behavior even in uniformly hyperbolic deterministic systems. This identifies hierarchy as a distinct structural mechanism capable of limiting the applicability of linear response theory in chaotic transport.

\bibliography{bibliography}
\end{document}